\documentclass[amssymb,prl,preprintnumbers,superscriptaddress,aps,nofootinbib,twocolumn,
showpacs]{revtex4-1}
\usepackage{graphicx, epsfig, amssymb} 
\usepackage{amsmath, amsfonts}
\usepackage{bm} 
\usepackage[breaklinks]{hyperref}
\usepackage[usenames,dvipsnames]{color}
\usepackage{enumerate}
\usepackage{dsfont}
\usepackage{xcolor}
\usepackage[normalem]{ulem}

\def\be{\begin{equation}}
\def\ee{\end{equation}}
\def\beq{\begin{eqnarray}}
\def\eeq{\end{eqnarray}}

\def\m{\mu}
\def\n{\nu}
\def\di{{\rm d}}

\def\DMa{DMa}
\def\DiM{DM{} }

\def\aj{Astronomical Journal }                 
\def\apjl{Astrophysical Journal }             

\def\apjs{ApJS }              

\def\aap{A\&A }  	

\newcommand{\nw}[1]{{}}

\begin{document}

\preprint{KCL-PH-TH/2019-14}

\title{Constraints on millicharged dark matter and axion-like particles\\
 from timing of radio waves}

\author{Andrea Caputo}  \affiliation{Instituto de F\'{i}sica Corpuscular, Universidad de 
Valencia and CSIC, Edificio Institutos Investigaci\'{o}n, Catedr\'{a}tico Jose Beltr\'{a}n 
2, Paterna, 46980 Spain }
\author{Laura Sberna} \affiliation{Perimeter Institute, 31 Caroline St N, Ontario, Canada}
\author{Miguel Fr\'ias}  \affiliation{Facultat de F\'{i}sica, Universitat de Barcelona, 
Mart\'{i} Franqu\`{e}s 1, 08028 Barcelona, Catalonia, Spain}
\author{Diego Blas}
  \affiliation{ Theoretical Particle Physics and Cosmology Group, Department of Physics,\\
   King's College London, Strand, London WC2R 2LS, UK}
\author{Paolo Pani}
  \affiliation{Dipartimento di Fisica, ``Sapienza'' Universit\`a di Roma \& Sezione INFN 
Roma1, Piazzale Aldo Moro 5, 00185, Roma, Italy.}
\author{Lijing Shao}   \affiliation{Kavli Institute for Astronomy and Astrophysics, Peking 
University, Beijing 100871, China}
 \author{Wenming Yan}
 \affiliation{Xinjiang Astronomical Observatory, CAS, 150 Science 1-Street, Urumqi, 
Xinjiang, 830011, China}

\begin{abstract} 
We derive constraints on millicharged dark matter and axion-like 
particles using pulsar timing and fast radio burst observations.  For dark matter particles of charge $\epsilon 
e$, the constraint from time of arrival (TOA) of waves is  ${\epsilon}/{m_{\rm milli}} \lesssim 10^{-8}{\rm eV}^{-1}$, for 
masses $m_{\rm milli}\gtrsim 10^{-6}\,$eV.  For axion-like particles,  the polarization of the
signals from pulsars yields a bound in the axial coupling  
$g/m_a\lesssim 10^{-13} {\rm GeV}^{-1}/(10^{-22}{\rm eV})$, for $m_a \lesssim 10^{-19}\,$eV. Both bounds scale as 
$(\rho/\rho_{\rm dm})^{1/2}$ for fractions
 of the total dark matter energy density $\rho_{\rm dm}$. We make a precise study of these bounds using TOA from several pulsars, FRB\,121102 and polarization measurements of PSR~J0437$-$4715. Our results rule out a new region of the 
parameter space for these dark matter models.
\end{abstract}


\maketitle


Unraveling the nature of dark matter~(\DMa) is among the most urgent issues in
fundamental physics. Indirect searches aim at detecting the effects of \DMa~in 
astrophysical observations, beyond its pure gravitational interaction. Given the feeble 
interaction of \DMa~with standard model fields, precise
measurements are particularly promising for these 
searches. 
When one requires precision, a particular measurement stands out in astrophysics: the 
time of arrival (TOA) of radio waves from pulsars and fast radio bursts (FRBs). The use 
of pulsar timing has already been suggested to study the effects of dark matter 
\cite{Khmelnitsky:2013lxt,Porayko:2014rfa,Pani:2015qhr,Clark:2015sha, Blas:2016ddr, 
Schutz:2016khr,DeMartino:2017qsa,Caputo:2017zqh,Dror:2019twh}.
In this  {\it Letter} we present new results for \DMa~models directly coupled to light 
from the propagation of radio pulses from pulsars and FRBs. A more comprehensive 
exploration will be presented elsewhere~\cite{followup}.

If \DMa~is coupled to the electromagnetic field, one expects modifications in the 
{\it emission}, {\it propagation}, and {\it detection} of radio pulses. We focus here on the effects during 
the propagation, which are robust under astrophysical uncertainties.
In particular, we derive stringent constraints on millicharged \DMa~and axion-like 
particles~(ALPs) based on dispersion measurements (DM) of radio signals from pulsars and 
FRBs, and on the modulation of the light polarization angle due to axion-like \DMa~in the 
Milky Way.

We give a unified treatment,  where the millicharged \DMa~and ALPs are considered as 
independent species.
In the former case we consider that (a fraction of) the \DMa~is made of particles 
with mass $m_{\rm milli}$ and electric charge $q= \epsilon e$ 
($\epsilon\ll1$)~\cite{DeRujula:1989fe,Perl:1997nd,Holdom:1985ag,Sigurdson:2004zp,
Davidson:2000hf,McDermott:2010pa,Berlin:2018bsc, Ejlli:2017uli}. {As an example, this coupling arises in models where the DMa is charged under a dark photon, which is kinematically coupled to the visible photon \cite{McDermott:2010pa,Berlin:2018bsc}. In our analysis we remain agnostic to  the origin of this term and other possible model-dependent signatures behind the charge of the DMa, and focus on constraining $\epsilon$.}
Regarding ALPs, 
we assume the existence of axion-like~\cite{Peccei:1977hh, 
Weinberg:1965nx, Preskill:1982cy, Abbott:1982af, Dine:1982ah}, pseudo-scalar DMa of mass $m_a$ (represented by the field $\phi$ below).

The relevant field equations read
\begin{eqnarray}
 (\square -m_a^2)\phi &=& -\frac{g}{4} F_{\mu\nu}\tilde F^{\mu\nu}\,,  \label{axion}\\
 \partial_\mu F^{\mu\nu}&=& 4\pi e j^\nu + 4\pi \epsilon e 
j^\nu_{\rm milli} -\frac{g}{2} 
\epsilon^{\mu\rho\lambda\nu} F_{\mu\rho}\partial_\lambda \phi \,,\label{Maxwell}
\end{eqnarray}
where $g$ is the ALP-photon axial coupling, $j^\nu$ is the ordinary electron current, whereas 
$j^\nu_{\rm milli}$ is the current from millicharged particles. {The role of this term in the  propagation of radio-waves will be 
studied in the next section, under the assumption of a cold distribution of the millicharge \DMa~ component.}

\vskip2pt
\noindent{\bf{\em Dispersion in the TOA.}}
We consider the propagation of a light signal of frequency $\nu=\omega/(2\pi)$ along the 
$z$ direction in the presence of a homogeneous background magnetic field polarized 
along (say) the $y$ direction, $\vec B=(0,B,0)$. We neglect a possible $B_z$ component in this section since its role in dispersion of the light signal in a medium of particles of mass $m_q$ and charge $q$ is suppressed by $ q B_z/(m_q \omega)$,  always small for the cases we study.
For the first part of this work, DMa is considered as a cold-medium with vanishing 
background values for the fields appearing in~\eqref{axion}-\eqref{Maxwell}. 
When $\omega\gg m_a$, the propagation 
of the light signal in this medium is described by the first-order system $ 
i\frac{\partial}{\partial z}\left|\psi(z)\right\rangle={\cal 
M}\left|\psi(z)\right\rangle$, where the $\left|\psi(z)\right\rangle$ is a 
linear combination of the two photon polarizations along the $x$ and $y$ directions and 
of the ALP state~\cite{Raffelt:1987im}. The 
$3\times3$ mixing matrix reads~\cite{Dupays:2005xs}
\begin{equation}
{\cal M}:= \left(\begin{array}{ccc}
        \omega+\Delta_{xx} & 0 & 0 \\
        0& \omega+\Delta_{yy} & g B/2 \\
        0 &  g B/2	& \omega-m_a^2/(2\omega) 
       \end{array}
\right)\,. \label{eq:M}
\end{equation}
%
 The
terms 
$\Delta_{xx}$ and $\Delta_{yy}$ contain both QED vacuum 
polarization effects and plasma effects~\cite{Raffelt:1987im,Dupays:2005xs}. 
The first ones are of order $ \Delta_{xx}^{\rm QED}\sim \Delta_{yy}^{\rm 
QED}\sim\omega\frac{e^2}{45\pi}\left(\frac{B}{B_c}\right)^2$,
where $B_c\approx 4\times 10^{13} \,{\rm 
G}$~\cite{Adler:1971wn}. We shall only consider interstellar magnetic fields, for which 
$B\ll B_c$ and $\Delta^{\rm QED}$  effects are negligible. Plasma effects arise from the 
presence of free charges. In the limit where the photon energy is much smaller than 
the mass of the charged, cold particles~\cite{GellMann:1954db,GOL64,Latimer:2013rja},  
\begin{equation}
 \Delta_{xx}^{\rm plasma}\sim \Delta_{yy}^{\rm plasma}\sim- 
\frac{\omega^2_p}{2\omega}\,, \label{eq:plasmaf}
\end{equation}
where $\omega_p^2:=\sum_i\frac{4\pi n_i q_i^2}{m_i}$ is the plasma frequency for particles with 
charge $q_i$, mass $m_i$, and number density $n_i$. The normal modes corresponding to 
\eqref{eq:M} satisfy
 \begin{eqnarray}
 k_0&=& \omega -\frac{\omega^2_p}{2\omega}\,,~~~~~
 k_\pm=\frac{4 \omega ^2-\omega_p^2-m_a^2 \mp \sqrt{\Delta_\omega}}{4 \omega 
}\,,\label{dispP}
\end{eqnarray}
with $\Delta_\omega=(m_a^2-\omega_p^2)^2+4 B^2 g^2 \omega^2$.
%
The last term in $\Delta_\omega$ is always subdominant and we treat it perturbatively.  

The TOA of a signal traveling at  speed $v=\partial \omega/\partial k$ across a 
distance $d$ is $T=\int_0^d \frac{\di l}{v}=\int_0^d \di l\frac{\partial k}{\partial 
\omega}$ along the line of sight. From the previous expressions one finds for the
relevant polarizations,
\begin{eqnarray}
 v _0^{-1}&=& 1+\frac{\omega^2_p}{2\omega^2}\,,\label{dispersionrelI} \\
 v_{-}^{-1} &=& v _0^{-1}+ \frac{B^2 g^2}{2 (m_a^2 -  \omega_p^2)}-  \frac{3B^4 g^4 \omega^2}{2(m^2_a - \omega_p^2)^3}. \label{dispersionrel}
\end{eqnarray}


In the absence of {new physics} $(\epsilon=g=0$),  the previous modes propagate with velocity $v_0$.
For a photon with frequency $\nu$, a background of 
cold free electrons yields a time delay 
\begin{align}
	\Delta t_{\rm \DiM}^{\rm astro} &= {\frac{1}{2\pi} \frac{e^2}{m_e}  \, \text{DM}_{\rm astro}  \, \left(\nu^{-2} - \nu_\infty^{-2} \right) }\label{DtDMastro} \\
	&\sim 4.15 \,\left(\frac{{\rm \DiM}_{\rm 
astro}}{{\rm 
	    pc} \,
{\rm cm}^{-3}}\right)\left(\frac{\nu}{{\rm GHz}}\right)^{-2} \,{\rm ms}\,,  \nonumber
\end{align}
relative to a photon with high enough energy {($\nu_\infty$ in the previous formula)} \cite{lorimer2005handbook}. Here ${\rm \DiM}_{\rm astro}:= \int n_e \di l $ is the standard 
dispersion measure~(DM) from electrons with number density $n_e$ along the light of 
sight. {The last line is also the observational definition of the dispersion measure, $ {\rm DM}_{\rm obs} $. }Comparing this number with the ALP-photon coupling term in 
Eq.~\eqref{dispersionrel} one sees that the modifications from the interstellar or 
intergalactic magnetic fields ($B\lesssim \m$G) are only relevant  for $g >\rm GeV^{-1}$, which is already excluded by other methods, e.g. \cite{Anastassopoulos:2017ftl}. We ignore these terms in the following. {We have checked that the high magnetic field of the pulsar magnetosphere is also not relevant for our studies and we ignore it. Finally, the local conditions of FRBs are not known. It is rather unlikely that they play a role in the DM and even more that they cancel the effects from the \DMa~plasma, Eqs.~\eqref{dispersionrelI} and \eqref{dispersionrel}. We hence restrict our analysis of the TOA to the millicharged \DMa. }

\noindent{\bf{\em TOA constraints on millicharged \DMa.}}
%
As we explained above, we now focus on the case of millicharged \DMa, i.e.  $g=0$. The contribution of the millicharged \DMa~to the time delay is given by an an expression analogue to \eqref{DtDMastro}, now considering the \DMa~ particles as the dispersive medium
\begin{equation}
{\Delta t_{\rm \DiM}^{\rm milli}= \frac{1}{2\pi} \frac{\epsilon^2 e^2}{m_{\rm milli}}  \, \int  \di l\,  n_{\rm milli}   \, \left(\nu^{-2} - \nu_\infty^{-2} \right) }. \label{DtDMmilli}
\end{equation}
In this case
the observed \DiM {is {dominated }by the sum of the contributions from 
ordinary electrons and millicharged particles (see also \cite{Gardner:2009et}), ${\rm 
\DiM}_{\rm obs} = {\rm \DiM}_{\rm astro} + {\rm \DiM}_{\rm 
milli}$, {where the millicharged contribution is obtained by comparing \eqref{DtDMastro} and \eqref{DtDMmilli},}
\begin{equation}
	{\rm \DiM}_{\rm milli} = \Big(\frac{\epsilon}{m_{\rm milli}}\Big)^2m_e\int \di l\,
	\rho_{\rm milli} \,,
\end{equation}
where $\rho_{\rm milli}$ is the density of millicharged particles, which is equal to 
or smaller than the full \DMa~density $\rho_{\rm dm}$.
While the effect of ${\rm \DiM}_{\rm astro}$ and ${\rm \DiM}_{\rm milli}$ are 
completely degenerate, for a source at a distance $d$ any measurement of the \DiM
can be translated 
into a \emph{conservative} upper bound on ${\epsilon}/{m_{\rm milli}}$ by simply 
requiring that all the \DiM~is due to \DMa, i.e.
${\rm \DiM}_{\rm milli}<{\rm \DiM}_{\rm obs}$. This yields
\begin{equation}
 \frac{\epsilon}{m_{\rm milli}} \lesssim \frac{10^{-8}}{\rm eV} \sqrt{\frac{0.3\,{\rm 
GeV}/{\rm cm}^3}{\rho_{\rm milli}}} \sqrt{\frac{{\rm 
\DiM}_{\rm obs}}{20\, 
{\rm 
pc}/{\rm cm}^{3}}} \sqrt{\frac{400\,{\rm pc}}{d}}\, \,, \label{pback}
\end{equation}
where we normalized the quantities by typical values within the galaxy.
This estimate gives already a rather stringent bound, which can be 
refined through a Bayesian analysis. In the following we closely 
follow~\cite{Shao:2017tuu}. Given our theoretical hypothesis (${\rm \DiM}_{\rm obs} = {\rm 
\DiM}_{\rm astro} 
+ 
{\rm \DiM}_{\rm milli}$), and the set of measurements of ${\rm \DiM}_{\rm obs}$ from 
$N$ pulsars, we construct the log-likelihood as
\begin{equation}
	\ln \mathcal{L} = -\frac{1}{2}\sum_{i=1}^N \frac{\Big({\rm \DiM}^i_{\rm obs} - 
{\rm \DiM}^i_{\rm astro}-{\rm \DiM}^i_{\rm milli}\Big)^2}{\sigma_i^2} \, .
\end{equation}
Here $\sigma_i$ is the dispersion for each pulsar, obtained adding in quadrature 
statistical uncertainties on ${\rm \DiM}_{\rm obs}^i$ and the astrophysical ones on ${\rm 
DM}_{\rm astro}^i$. 
We used a uniform prior on $\epsilon/m_{\rm milli}>0$ and verified that our results do 
not depend on this choice.

We shall consider two datasets of pulsars extracted from the 
ATNF Pulsar Catalogue~\cite{2005AJ....129.1993M} as explained in the 
Supplement Material. In both cases we assume a Navarro-Frenk-White profile for the \DMa~density, normalized to {a local value of}
$\rho_{\rm dm} \approx 0.3\,{\rm GeV}/{\rm cm}^3$. 
The first dataset comprises $N=13$ local pulsars  with the smallest values of ${\rm \DiM _ 
{\rm obs}}/d $ and for which parallax measurements of the distance $d$ are 
available. We only choose pulsars located away from the galactic plane. This is to 
minimize the effect of the evacuation of DMa from the galactic plane for 
millicharged~\DMa. While early studies argue that this effect is relevant for  
$\epsilon\gtrsim 5.4\times 10^{-22}\Big(\frac{m_{\rm milli}}{\rm 
eV}\Big)$~\cite{McDermott:2010pa,Chuzhoy:2008zy}, a recent study~\cite{Dunsky:2018mqs} 
suggests that this bound may be too restrictive. 
We also consider a second dataset of $N_{\rm cluster}=13$ pulsars located in globular 
clusters within $8\,{\rm kpc}$ from the galactic center and off the disk, again with the 
smallest ${\rm \DiM _ {\rm obs}}/d $. Distances of clusters can be determined by 
different methods~\cite{Krauss65} not relying on the DM, and their uncertainty is usually 
of a few percent. We therefore assign a conservative error of $10\%$ to the value of $d$ 
for the pulsars in this second dataset. Even if the effect of the galactic magnetic 
field on the density of millicharged \DMa~away from the galactic disk is uncertain, 
we do not expect \DMa~to be evacuated at high galactic latitudes, and our analysis should 
provide realistic constraints.

For each pulsar we compute ${\rm \DiM}_{\rm astro}^i \approx \langle n_e\rangle_i d_i$,
where $\langle n_e\rangle_i$ is an average electron density along the line of sight 
obtained using the YMW16 model~\cite{Yao}, while $d_i$ is the pulsar distance obtained
from parallax (for the first dataset) or from the location of the globular cluster 
(for the second dataset). In the former case, we assign $\langle n_e\rangle_i$ a $ 20 \% 
$ error to take into account potential systematics in the electron density model. This is a conservative 
approach given the uncertainties in~\cite{Yao}.
We perform a Monte-Carlo Markov chain analysis using the {\scshape Python} ensemble 
sampler {\scshape Emcee} 
\cite{ForemanMackey:2012ig} to explore the posterior distribution. For our datasets, 
$10^5$ 
samples are accumulated with $20$ chains. The chains show good acceptance rate and 
convergence.
The results are similar for the two datasets:
\begin{equation}
\frac{\epsilon}{m_{\rm milli}} \lesssim \frac{4 \times 10^{-9}}{\rm eV}\sqrt{\frac{0.3\,{\rm 
GeV}/{\rm cm}^3}{\rho_{\rm milli}}} ~\,\,{\rm at}\,\, 
95\%~{\rm C.L.} \label{eqDISp}
\end{equation}
which we compare to other existing bounds in Fig.~\ref{fig:epsilonm}. In particular these results 
are compatible with $\epsilon=0$. For completeness, we 
also show a similar (weaker) bound estimated from the dispersion of the fast radio burst 
FRB\,121102~\cite{Chatterjee:2017dqg}. This line falls in the ballpark of the estimate \eqref{pback}. A more comprehensive analysis for FRBs will be 
presented elsewhere~\cite{followup}.

The mass range in Fig.~\ref{fig:epsilonm} is limited on the left because the 
expression~\eqref{eq:plasmaf} is valid as long as the energy of the photon is {smaller} than 
$m_{\rm milli}$. For radio waves from pulsars, $m_{\rm milli} \gtrsim \omega\sim {\rm GHz} 
\sim 10^{-6}~{\rm eV}$. Since the bound is more stringent for small masses, these 
constraints could improve {as $1/m_{\rm milli}$} for sub-GHz pulsar measurements  {in systems with properties similar to the ones used in our analysis. Low-frequency measurements are indeed possible, see 
e.g. Ref.~\cite{Pilia:2015tqa}, though we leave a more systematic study of the sources for the future.}
Figure~\ref{fig:epsilonm} shows that our bounds are competitive for masses below the 
Tremaine-Gunn bound on fermionic \DMa, $m_{TG}\gtrsim \rm keV$~\cite{Tremaine:1979we}. 
Hence, they apply to scalar charged \DMa~or to models with a fraction of millicharged 
fermionic \DMa~(see Eq.~\eqref{eqDISp} for the scaling of the bound with $\rho_{\rm 
milli}$).

{Finally, the existence of milli-charge \DMa~  also impacts the cosmological 21-cm line and distortions of the CMB 
\cite{Ali-Haimoud:2015pwa,Munoz:2018pzp,Slatyer:2018aqg}. It seems possible that these observations also constrain the 
very light case considered here, though previous studies focus on much heavier \DMa~candidates, and it seems cautious 
not to extrapolate their conclusions at much lower masses. Instead, it would be interesting to extend these analyses to 
smaller masses in the future.}

\begin{figure}[th]
\centering
\includegraphics[width=1.05\linewidth]{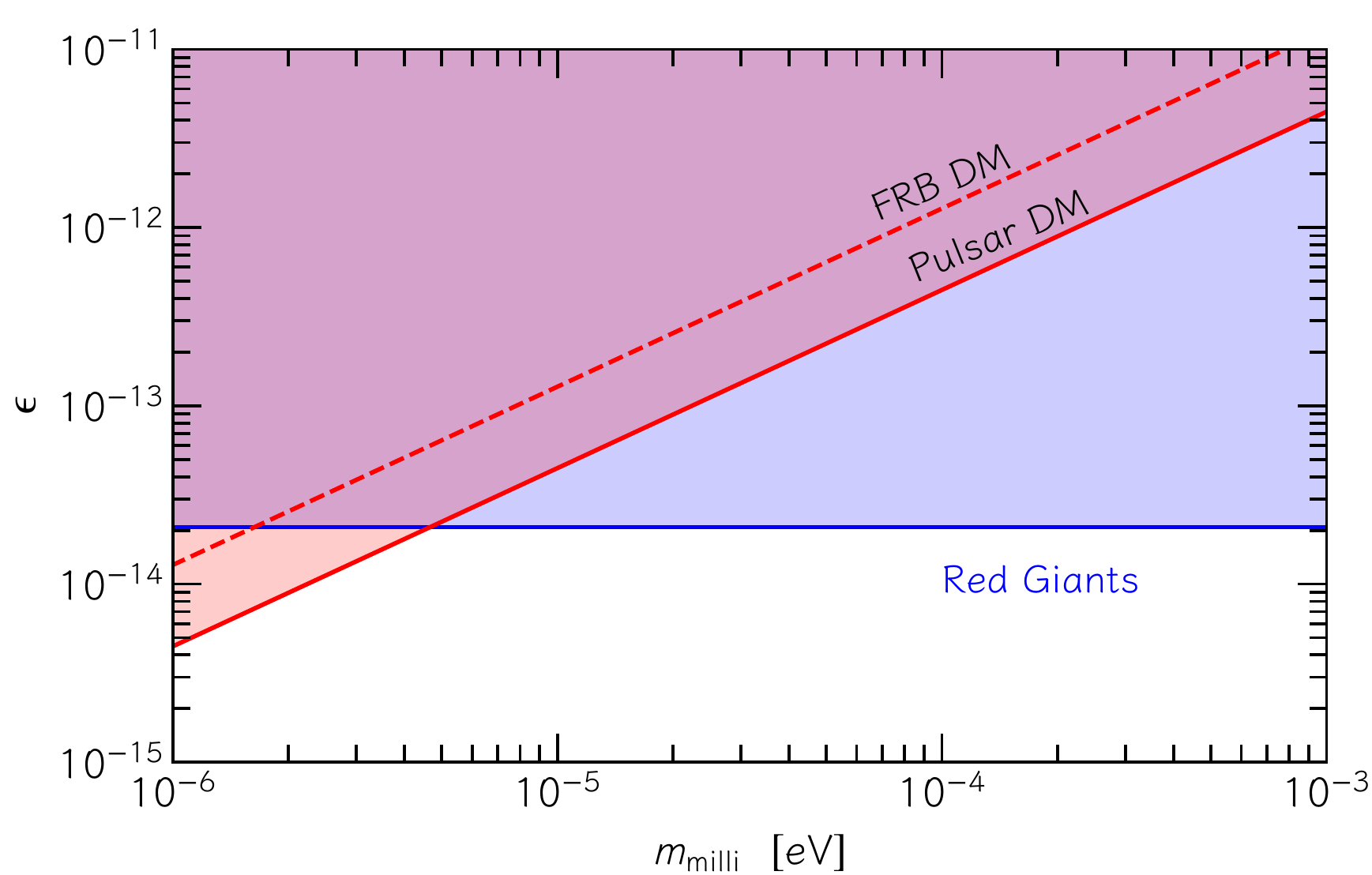}
  \vspace{-.1cm}
\caption{\em 
Constraint on millicharged \DMa~in the $\epsilon-m_{\rm milli}$ space from pulsar 
(solid red line) and FRB\,121102 (dashed red line) \DiM~at 
$95\%$ confidence level. Solid blue line indicates the bound from Red 
Giants~\cite{Davidson:2000hf}. We assume a homogeneous \DMa~density $\rho_{\rm 
dm}=\rho_{\rm milli} \approx 0.3\,{\rm GeV}/{\rm cm}^3$.
The bound scales as $\rho_{\rm milli}^{-1/2}$ for 
fractional components.
} \label{fig:epsilonm}
\end{figure}

\begin{figure}[!htb!]
\centering
\includegraphics[width=1.12\linewidth]{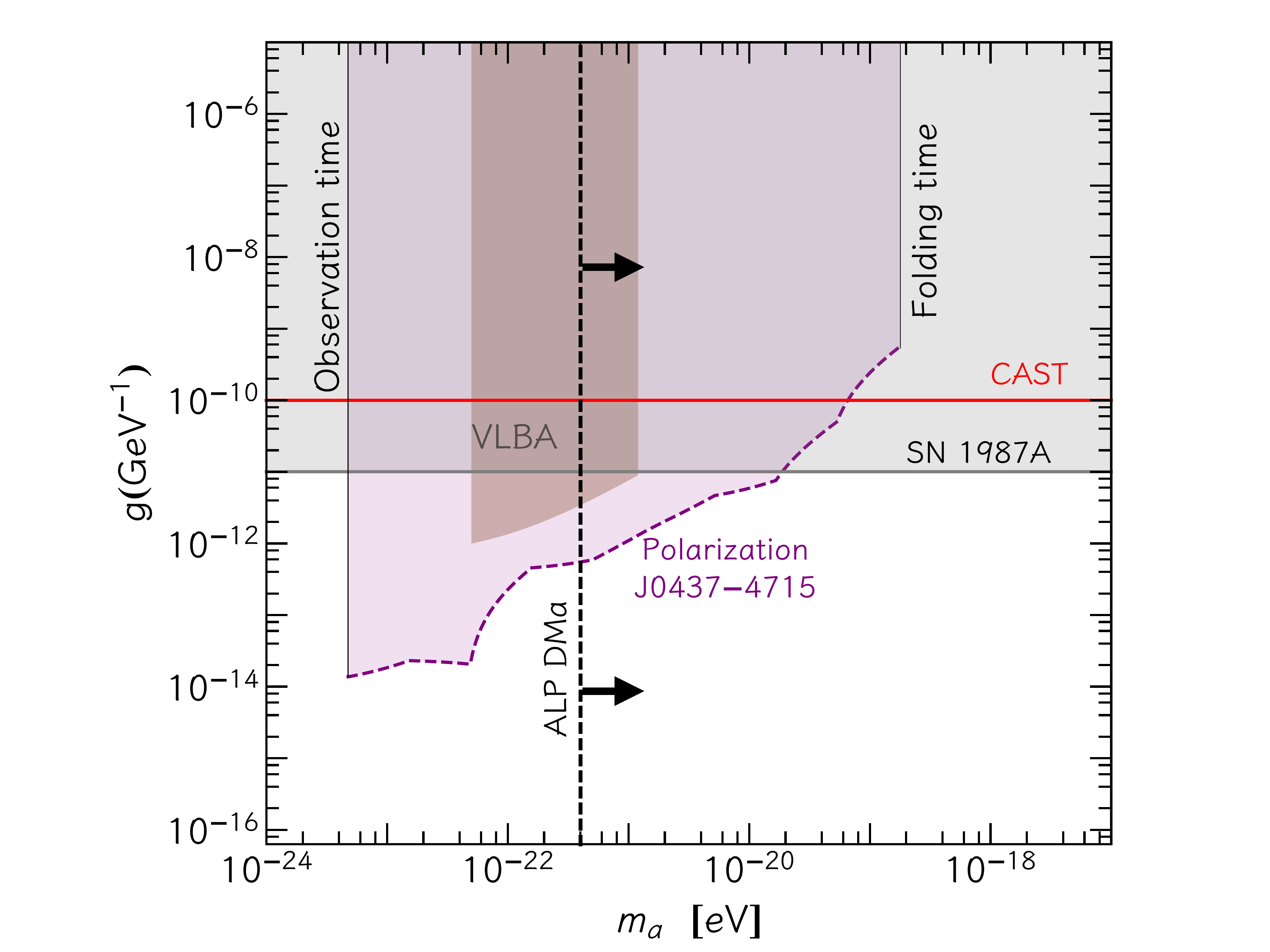}
  \vspace{-.1cm}
\caption{\label{fig:ALPs}\em
Constraints for ALP \DMa~in the plane $g-m_a$ at $95 \%$ CL. The dash-dotted purple line indicates the 
lower bound set by polarization measurements using real data. The darker gray band 
indicates the region excluded by CAST experiment~\cite{Anastassopoulos:2017ftl} and by 
supernova cooling~\cite{Payez:2014xsa}, while the amaranth pink area indicates the region 
excluded by MOJAVE VLBA polarization observations of parsec-scale jets from active 
galaxies~\cite{Ivanov:2018byi}. The vertical dashed line represents an estimation of the masses for which the ALP DMa candidate can constitute all the $\rho_{\rm dm}$ \cite{Kobayashi:2017jcf,Bar:2018acw,Marsh:2018zyw}.}
\label{TOAaxion}
\end{figure}

\noindent{\bf{\em Polarization constraints on ALPs.}}
%
We now consider the case where the millicharged particles are absent, $j^\n_{\rm milli}=0$.
As discussed before, the modification of the TOA from the terms depending on $g$ in 
Eq.~\eqref{dispersionrel} is negligible and we ignore it. 
Nevertheless, due to their pseudo-scalar nature, ALPs also induce an oscillating variation 
of light polarization~\cite{Harari:1992ea, Ivanov:2018byi, Sigl:2018fba, 
Fujita:2018zaj, Plascencia:2017kca, Obata:2018vvr, Ejlli:2016avx}. Parity-symmetry breaking leads to birefringence, i.e. 
different phase velocities for left- and right-handed modes, which in turn induces 
rotation of the 
linear polarization plane. At first approximation,  we assume the  ALP-DMa background in the Milky Way rest frame to be described by the field configuration \cite{Arvanitaki:2017nhi}
\begin{equation}
\phi(x,t)=\tilde\phi_0(x) \int \di^3 v\, e^{-\frac{v^2}{\sigma^2_0}} e^{i (\omega_v t- m_a 
\vec{v}\cdot\vec{x})+i\varphi_{\bf v}}+c.c., \label{eq:field}
\end{equation}
where $\sigma_0\approx 10^{-3}$ corresponds to the virialized velocity of the Milky Way and $\varphi_{\bf v}$ are arbitrary phases. The value $\tilde\phi_0$  changes smoothly with $x$ to reproduce the DMa  energy density.  Finally, for this non-relativistic configuration one can assume that $\omega_v\approx m_a(1+v^2/2)$.
 For low DMa masses, this field configuration has only long modes as compared to the wavelength of radio signals and an eikonal approximation can be used to study the propagations of waves in this continous background \cite{PhysRev.126.1899}. The leading result of this calculation yields an effect for the polarization angle of a photon 
propagating from time $t$ to $t+T$  \cite{Ivanov:2018byi,Harari:1992ea} 
%
%
%
\be
\begin{split}
	\theta(t,T)\sim &1.4\times 10^{-2}\sin(m_a t +\delta)\\
	&\left(\frac{g}{10^{-12}\,{\rm GeV}^{-1}}\right)\frac{10^{-22}\rm \, eV}{m_a}\ \rm rad,
	\label{polangle}
\end{split}
\ee
where $\delta$ is a phase over which we will marginalize.
The characteristic time scale for the axion background oscillation is $T_{\rm ALP}\sim 
\frac{10^{-22} {\rm eV} }{m_a} \,  {\rm yr}$; if one continuously observes the polarized 
light from 
the source during a time $t_{\rm obs} \gtrsim T_{\rm ALP}$, 
the observed variation of the polarization angle~\eqref{polangle} may  
constrain the amplitude of the axion oscillations\footnote{{Notice that after exploring a quarter of a period of oscillation, the original value of $\delta$ is not relevant. Hence, even if a system lives in a region with $\delta\ll1$, the previous analysis is valid for masses satisfying  $t_{\rm obs} \gtrsim T_{\rm ALP}$}.} , i.e. the coupling $g$
for a given mass $m_a$. 
Pulsars are observed for long periods and the polarization angle is measured to be almost 
constant with a precision of roughly one degree, that can be compared with
Eq.~\eqref{polangle}. We use the polarization data from Ref.~\cite{Yan:2011bq} and 
in particular PSR~J0437$-$4715, which is the pulsar with the highest number of 
observations of the polarization angle, spanning a period of roughly four years. The 
ionospheric contribution to the polarization angle was subtracted using the program 
{\scshape Getrm-Iono}~\cite{Han:2006ci}. Similar results are obtained when the ionospheric contribution is subtracted with the alternative
FARROT method  developed at the Dominion Radio Astrophysical Observatory (DRAO), Penticton, Canada.
We performed a likelihood estimation of the coupling $g$ for
a set of fixed masses $m_a$. For each value of the mass, we marginalize over the unknown 
phase $\delta$ in Eq.~\eqref{polangle} in the interval $[-\pi,\pi]$
and then obtained the $95\%$ C.L. exclusion value for $g$, which is our reported
constraint.  There is a caveat in using the bound from a single system: it may be that the pulsar of interest lives in a region where the amplitude of the field \eqref{eq:field} is lower than expected from the NFW profile. This situation may happen, for instance, in certain ULDM models where the field $\phi$ interpolates between different domains of condensation. The chances for this to happen are slim. Still, it is important to take this caveat into consideration. The use of more pulsars in the future will likely reduce this possibility even more.

The excluded region in Fig.~\ref{TOAaxion} spans roughly four orders of magnitude in the 
mass range, from  $m_a\sim 10^{-19}\,{\rm eV}$ to $m_a \sim 10^{-23}\,{\rm eV}$. The lower limit 
is set by the total observation time ($\sim 4\,{\rm yr}$), whereas the upper limit
is set by the resolution time in the data-set during each observation run (`folding time'), that is roughly 1 hour for J0437$-$4715.
The derived lower bounds scale as
$1/m_a$ ---~with some 
modulation due to the fact that observations of the polarization angle for J0437$-$4715 
are not homogeneous in time~--- and are stronger for smaller masses, i.e longer observation 
time. The bound scales as $\sim \sqrt{\rho_{\rm dm}}$, so it can be 
competitive even if ALPs form only a small fraction of the \DMa.  
This is particularly important at low masses, where other astrophysical 
constraints require the mass of the  ALP  to be ~$m_a\gtrsim
10^{-21}\,$eV if it constitutes all the \DMa. 
These bounds are based on the clustering properties of the \DMa{}  candidate at small scales ~\cite{Kobayashi:2017jcf}, 
the modifications of rotation curves in the inner regions of galaxies \cite{Bar:2018acw}, and the mere existence of 
galaxies with very small gravitational binding energies~\cite{Marsh:2018zyw}.
These constraints are subject to independent astrophysical uncertainties, though together they indicate that masses 
below $m_a\sim10^{-21}\,$eV are in tension with current data. In Fig.~\ref{TOAaxion} we represent the previous limit 
by a conservative line at $m_a=4\times10^{-21}\,$eV. This limitation relaxes for fractional components.

\noindent{\bf{\em Discussion.}}
Several \DMa~models introduce dispersion effects in the photon propagation. 
Although small, these effects accumulate for photons coming from astrophysical sources and 
can be constrained through precision measurements. 
The effect of millicharged \DMa~is degenerate with that of ordinary plasma and 
improving models for the local plasma distribution will help strengthening the 
constraints from DM. On the other hand, the effect of ALP-photon coupling is more striking 
and requires a careful analysis of the TOA as a function of the frequency.
In addition, in the upcoming era of the Square Kilometre Array, we will benefit
from a much larger pulsar sample (possibly comprising sources near the galactic center, 
where the \DMa~density is higher than what assumed here), combined with a significantly 
improved timing precision~\cite{Kramer:2015jsa, Shao:2014wja, Bull:2018lat}. The prospects 
of using radio waves in probing DMa are very promising in the near future. For ALPs, 
their coupling to photons generates an oscillation of the polarization angle of photons in 
the ultra-light \DMa~case. Our results in Fig.~\ref{TOAaxion} show that, for the mass 
range $10^{-23}-10^{-20}$eV, the constraints derived here are the best available and will 
greatly improve in the future with more data.  

We have considered propagation in a weak magnetic field for which dispersion due to the 
ALP-photon coupling and QED vacuum polarization effects are negligible. However, 
our formalism can be easily extended to include such effects, which might be relevant for 
propagation in strongly magnetized regions. A discussion of this effect 
will appear elsewhere~\cite{followup}.
 
\noindent{\bf{\em Note:}} While this work was close to completion, Ref.~\cite{Liu:2019brz} 
appeared on the arXiv, estimating constraints on ALPs using the polarization angle of 
radio waves from pulsars similar to those derived in the second part of our work. Even 
though the idea is similar, our analysis, based on real data, is distinct and the results 
differ from the ones in \cite{Liu:2019brz} by roughly a 
factor $\sqrt{\frac{10^{-22}{\rm eV}}{m_a}}\sqrt{\frac{400{\rm pc}}{d}}$ originating from
a different assumption about the $\phi$ configuration. 

\noindent{\bf{\em Acknowledgments.}}
We are grateful to Nikita Blinov, Richard Brito, Anson Hook, Georg Raffelt, 
G\"{u}nter Sigl for interesting discussions. We thank Davide Racco and  Mikhail M.~Ivanov for pointing out a 
mistake in the first arXiv version of this work. 
AC acknowledge support from national grants FPA2014-57816-P, FPA2017-85985-P and the
European projects H2020-MSCAITN-2015//674896-ELUSIVES and H2020-MSCA-RISE2015.
PP acknowledge financial support provided under the European Union's H2020 ERC, Starting 
Grant agreement no.~DarkGRA--757480, and support from the Amaldi Research Center funded 
by the MIUR program ``Dipartimento di Eccellenza''~(CUP:~B81I18001170001) and by the 
GWverse COST Action~CA16104, ``Black holes, gravitational waves and fundamental 
physics.''
Research at Perimeter Institute is supported by the Government of Canada through Industry 
Canada and by the Province of Ontario through the Ministry of Research and Innovation.
LS was partially supported by the National Science Foundation of China
(11721303), and XDB23010200.


\appendix*

\section{Supplemental material}

We provide here additional details of the datasets analyzed in this work.
For the millicharged \DMa, we analyzed a first set of galactic pulsars selected for their 
minimal ${\rm DM}/d $, where $d$ is derived from parallax, and for their good agreement 
with the electron density model (Table~\ref{tab:pulsargal}). A second set of pulsars is 
selected in galactic clusters (Table~\ref{tab:pulsarclus}). In this case, in addition to 
the aforementioned criteria, we also require that the pulsars are not further from the 
galactic center than the Solar System, $ \sqrt{X^2+Y^2+Z^2} < 8.3 \, {\rm kpc} $, and are 
also located far from the galactic disk, $ |Z|> 1 \, {\rm kpc} $. 
\begin{table}[t!]
	\centering
	\begin{tabular} {  l   c   c   c   } 
		\hline
		Pulsar & Parallax $({\rm mas})$ &   DM $({\rm pc\,cm}^{-3})$   &   $ n_e \ ({\rm cm}^{-3}) $ \\	
		\hline
		J1024$-$0719 	& 0.770   $\pm$  0.23 &  6.4778   $\pm 0.0006$ &  0.009036 \\
		J1012+5307	& 0.710   $\pm$  0.17  &  9.02314  $\pm$ 0.00007&  0.007827 \\
		J2010$-$1323	& 0.300   $\pm$  0.10  &  22.177  $\pm$ 0.005&  0.004931 \\
		J2234+0611 	& 0.700   $\pm$  0.20  &  10.7645 $\pm$ 0.0015&  0.008292 \\
		J1909$-$3744 	& 0.810   $\pm$  0.03  &  10.3932 $\pm$ 0.01&  0.016935 \\
		B2020+28	& 0.370   $\pm$  0.12  &  24.63109 $\pm$ 0.00018&  0.012689 \\
		B1508+55	& 0.470   $\pm$  0.03  &  19.6191  $\pm$0.0003&  0.004691 \\
		J2017+0603 	& 0.400   $\pm$  0.20  &  23.92344 $\pm$0.00009&  0.011004 \\
		B1534+12	& 0.860   $\pm$  0.18  &  11.61944   $\pm$  0.00002&  0.009608 \\
		J0108$-$1431 	& 4.200   $\pm$  1.40  &  2.38   $\pm$  0.19&  0.009246 \\
		B0031$-$07    	& 0.930   $\pm$  0.08  &  10.922  $\pm$ 0.006&  0.008167 \\
		J1023+0038  	& 0.731   $\pm$  0.022  &  14.325  $\pm$ 0.01&  0.008209 \\ 
		B1237+25	& 1.160   $\pm$  0.08  &  9.25159 $\pm$ 0.00053&  0.008940 \\
		\hline
	\end{tabular}
	\caption{List of local pulsars considered in this work.}
	\label{tab:pulsargal}
\end{table}


\begin{table}[t!]
	\centering
	\begin{tabular} {  l  l  c  c  c  c } 
		\hline
		Pulsar & Cluster & $d(\rm pc)$  &   DM $({\rm pc\,cm}^{-3})$   &   $ n_e \ ({\rm cm}^{-3}) $ \\	
		\hline
		B1516+02B  & M5 & 8500   &  29.47  $\pm 0.11$ &  0.000027 \\
		B1516+02A	& M5 & 8500   &  30.08   $\pm$ 0.05 &  0.000027 \\
		J1518+0204D	& M5 & 8000   &  29.3   $\pm$ 0.11&  0.000042 \\
		J1518+0204E	& M5 & 8000   &  29.3   $\pm$ 0.11 &  0.000042 \\
		J1518+0204C	& M5 & 8000   &  29.3146   $\pm$ 0.006 &  0.000042 \\
		J2140-2310A	& M30 & 9200   &  25.0640  $\pm$ 0.0041  &  0.000015 \\		
		J2140-2310B & M30 & 9200   &  25.09  $\pm$ 0.12 &  0.000015 \\		
		J0024-7204X & 47Tuc & 4690   &  24.539   $\pm$ 0.005 &  0.0083 \\		
		J0024-7204Z & 47Tuc & 4690   &  24.47   $\pm$ 0.01 &  0.0083 \\	
		J0024-7204Z & 47Tuc & 4690   &  24.29   $\pm$ 0.03 & 0.0083 \\ 
		B0021-72H & 47Tuc & 4690   &  24.37  $\pm$  0.02 & 0.0083 \\ 
		B0021-72E & 47Tuc & 4690   &  24.236 $\pm$  0.002& 0.0083 \\ 
		J0024-7204R & 47Tuc & 4690   & 24.361  $\pm$  0.007 & 0.0083 \\ 		
		\hline
	\end{tabular}
	\caption{List of pulsars in globular clusters considered in this work.}
	\label{tab:pulsarclus}
\end{table}

\bibliography{biblio}
\end{document}